\documentclass{article}

\usepackage{arxiv}

\usepackage[utf8]{inputenc} 
\usepackage[T1]{fontenc}    
\usepackage{hyperref}       
\usepackage{url}            
\usepackage{booktabs}       
\usepackage{amsfonts}       
\usepackage{nicefrac}       
\usepackage{microtype}      
\usepackage{graphicx}
\usepackage{natbib}
\usepackage{doi}
\usepackage{multirow}
\usepackage{amsmath}
\usepackage{array} 
\newcolumntype{C}[1]{>{\centering\arraybackslash}p{#1}} 

\title{Enhancing Automatic Modulation Recognition for IoT Applications Using Transformers}

\author{
	{\hspace{1mm}Narges Rashvand} \\
	Department of Electrical and Computer Engineering\\
	The University of North Carolina at Charlotte\\
	Charlotte, NC, USA \\
	\texttt{nrashvan@uncc.edu} \\
 \And
	{\hspace{1mm}Kenneth Witham} \\
	Kostas Research Institute at Northeastern University\\
	Boston, MA, USA \\
	\texttt{k.witham@kri.neu.edu} \\
  \And
	{\hspace{1mm}Gabriel Maldonado} \\
	Department of Electrical and Computer Engineering\\
	The University of North Carolina at Charlotte\\
	Charlotte, NC, USA \\
	\texttt{gmaldon2@uncc.edu} \\
  \And
	{\hspace{1mm}Vinit Katariya} \\
	Department of Electrical and Computer Engineering\\
	The University of North Carolina at Charlotte\\
	Charlotte, NC, USA \\
	\texttt{vkatariy@uncc.edu} \\
  \And
	{\hspace{1mm}Nishanth Marer Prabhu} \\
	Dept. of Electrical and Computer Engineering\\
 Northeastern University\\
	Boston, MA, USA\\
	\texttt{marerprabhu.n@northeastern.edu} \\
  \And
	{\hspace{1mm}Gunar Schirner} \\
	Dept. of Electrical and Computer Engineering\\ Northeastern University\\
	Boston, MA, USA\\
	\texttt{schirner@ece.neu.edu} \\
 \And
	{\hspace{1mm}Hamed Tabkhi} \\
	Department of Electrical and Computer Engineering\\
	The University of North Carolina at Charlotte\\
	Charlotte, NC, USA  \\
	\texttt{htabkhiv@uncc.edu} \\
}

\begin{document}
\maketitle

\begin{abstract}
Automatic modulation recognition (AMR) is vital for accurately identifying modulation types within incoming signals, a critical task for optimizing operations within edge devices in IoT ecosystems. This paper presents an innovative approach that leverages Transformer networks, initially designed for natural language processing, to address the challenges of efficient AMR. Our transformer network architecture is designed with the mindset of real-time edge computing on IoT devices. Four tokenization techniques are proposed and explored for creating proper embeddings of RF signals, specifically focusing on overcoming the limitations related to the model size often encountered in IoT scenarios. Extensive experiments reveal that our proposed method outperformed advanced deep learning techniques, achieving the highest recognition accuracy. Notably, our model achieves an accuracy of 65.75 on the RML2016 and 65.80 on the CSPB.ML.2018+ dataset.
\end{abstract}


\section{Introduction}
In wireless communication systems, Internet of Things (IoT) devices acting as transmitters utilize various modulation techniques to optimize data transmission rates and bandwidth utilization. While the transmitters determine the modulation type, IoT devices acting as receivers often lack this information. Automatic Modulation Recognition (AMR) plays a crucial role in non-cooperative communication systems and IoT applications, serving as a fundamental component for demodulating unknown signals. Its significance extends to various applications including spectrum sensing, signal surveillance, interference localization, and cognitive radio \cite{9037268, zhou2022automatic}. 

Approaches to AMR are broadly classified into two main categories: the likelihood-based (LB) approach and the feature-based (FB) methods. While the LB method is capable of achieving remarkable results, it requires prior knowledge of the probability density function (PDF), adding considerable complexity to the process. In contrast, FB approaches concentrate on directly extracting features from the signal, eliminating the necessity for additional channel or signal information and thereby reducing computational demands. These methods depend significantly on extracting and analyzing signal features \cite {zhou2022automatic}. 

Deep learning (DL) has recently made notable advancements, demonstrating remarkable potential across a range of fields, from smart cities \cite{icores24, khaleghian2023electric} and computer vision\cite {vaswani2017attention, gholami2023federated} to signal processing \cite {o2018over, krzyston2020high}. Furthermore, its application in wireless communications, especially in DL-based AMR techniques, has been expanding. 
Recent advancements in neural network architectures for the AMR tasks have broadened from convolutional-based neural networks to include innovative designs like Residual Networks (ResNet) \cite{o2018over} and Densely Connected Convolutional Networks (DenseNets) \cite{krzyston2020high}.  LSTM-based recurrent neural networks (RNNs) \cite{rajendran2018deep} and convolutional LSTM deep neural networks (CLDNNs) \cite{ramjee2019fast}, have also been introduced in this domain. CLDNNs and ResNets are particularly effective at lower signal-to-noise ratios (SNRs), whereas LSTMs and ResNets are better at higher SNRs. However, none of these architectures demonstrate superior performance across all SNR values \cite{hamidi2021mcformer}. Moreover, due to their complex nature, these models may not be the most suitable choices for the constrained resources of IoT devices \cite {zhou2022automatic, usman2020amc}.

In recent years, the self-attention mechanism, introduced in \cite{vaswani2017attention}, has emerged as a significantly improved method for modeling sequences. This innovation, introduced through the Transformer model, addresses the parallelization challenges inherent in recurrent neural networks (RNNs) and has achieved broad adoption across the field of computer vision and natural language processing (NLP). The self-attention mechanism is the core of transformer architecture, evaluating the correlations between pairs of tokens (symbols or units within sequence data) across the sequence. Transformers have shown promising results in various sequential pattern recognition tasks and especially, the introduction of the Vision Transformer (ViT) model \cite{dosovitskiy2020image} marked a novel application of the Transformer's encoder for image classification tasks.

Despite the rapid advancement of Transformers in the computer vision field, their potential to recognize signal modulation patterns is still largely unexplored. Only a few studies in wireless communications and AMR have explored this area \cite{rajagopalan2023transformers,hamidi2021mcformer,9826820}. 
AMR involves processing sequential data, time-series signals, and Transformer networks have the high potential to handle time-series data via their self-attention mechanism. This mechanism enables them to capture dependencies across different parts of the input sequence, presenting a significant opportunity for exploration in the field of AMR. Additionally, Transformers can process input sequences in parallel, unlike RNNs, leading to faster training and inference, a crucial advantage for real-time applications in IoT environments. Furthermore, their scalability and ability to handle variable-length sequences make them well-suited for processing signals of varying duration.


Inspired by vision transformers, we introduce Transformer-based architectures for modulation recognition tasks. The proposed Transformer-based AMR approaches enable rapid processing and minimal resource usage, critical aspect of edge computing systems within IoT ecosystems. By integrating our proposed models into edge devices,  our research not only promises improvements in edge computing but also addresses key challenges such as bandwidth efficiency and energy consumption. This ensures that IoT devices can operate longer and more effectively within their constrained environments.
 
The rest of this paper is organized as follows. Section 2 addresses the two datasets, RadioML2016.10b and our modified CSPB.ML.2018 dataset with channels (CSPB.ML.2018+), and emphasizes their properties. Section 3 focuses on the transformer-based architecture for the task of AMR, followed by an introduction to four distinct tokenization strategies. Experiments and discussions are detailed in Section 4, with future investigation in Section 5 and conclusions summarized in Section 6.

\section{Datasets}

\begin{itemize}
\item RadioML2016.10b\cite{o2016radio}

This dataset is composed of ten modulations, including eight digital and two analog modulation types over SNR values ranging from -20 dB to +18 dB, in increments of 2 dB, i.e, $\{-20,-18,...,+16,+18\}$. These samples are uniformly distributed across this SNR range. The dataset, which includes a total of 1.2 million samples with a frame size of 128 complex samples, is labeled with both SNR values and modulation types. The dataset is split equally among all considered modulation types. At each SNR value, the dataset contains 60,000 samples, divided equally among the ten modulation types, with 6,000 samples for each type. For the channel model, simple multi-path fading with less than 5 paths were randomly simulated in this dataset. It also includes random channel effects and hardware-specific noises through a variety of models, including sample rate offset, noise model, center frequency offset, and fading model. Thermal noise was used to set the desired SNR of each data frame.
 
\item CSPB.ML.2018+\cite{spoonerDatasetMachineLearningChallenge}

This dataset is derived from the CSPB.ML.2018\cite{spoonerDatasetMachineLearningChallenge} dataset which aims to solve the known problems and errata \cite{AllBPSKSignals} with the RadioML2016.10b\cite{o2016radio} dataset. CSPB.ML.2018 only provides basic thermal noise as the transmission channel effects. We extend CSPB.ML.2018 by introducing realistic terrain-derived channel effects based on the 3GPP 5G channel model \cite{3gpp.38.901}. CSPB.ML.2018+ contains 8 different digital modulation modes, totaling 3,584,000 signal samples. Each modulation type has signals with a length of 1024 I/Q samples. Channel effects applied include slow and fast multi-path fading, Doppler, and path loss. The transmitter and receiver placements for the 3GPP 5G channel model are randomly selected inside a 6x6 km square. The resulting dataset covers an SNR ($E_s/N_0$) range of -20 to 40dB with the majority of SNRs distributed log-normally with $\mu_{SNR}\approx0.71dB$ and $\sigma_{SNR}\approx8.81dB$ using $SNR_{dB}=10\log_{10}(SNR_{linear})$ as the log conversion method.

\end{itemize}

\noindent The characteristics of these two datasets are detailed in Table \ref{tab_dataset}.

\begin{table}[htpb]
\centering
\caption{Dataset Characteristics: RadioML2016.10b \cite{o2016radio} vs. CSPB.ML.2018+\cite{spoonerDatasetMachineLearningChallenge} }
\label{tab_dataset}
\resizebox{\columnwidth}{!}{%
\begin{tabular}{|c|p{6cm}|p{6cm}|}
\toprule[\heavyrulewidth] \midrule
&  \textbf{RadioML2016.10b \cite{o2016radio}}   & \begin{tabular}[c]{@{}c@{}}\textbf{CSPB.ML.2018+\cite{spoonerDatasetMachineLearningChallenge}}\\ \end{tabular} \\ \midrule
\multirow{1}{*}{\textbf{Number of Modulation types}} 
&  10 (8 digital and 2 analog modulations) &  8 digital modulations        \\

\hline
\multirow{2}{*}{\textbf{Modulation pool}}
&  BPSK, QPSK, 8PSK, QAM16, QAM64, BFSK, CPFSK, PAM4, WBFM, AM-DSB
&  BPSK, QPSK, 8PSK, DQPSK, MSK, 16-QAM, 64-QAM, 256-QAM  \\

\hline
\multirow{1}{*}{\textbf{Signal length}} 
&   128&  1024       \\ 
\hline
\multirow{1}{*}{\textbf{SNR range}} 
& -20 dB to +18 dB&    -19 dB to +40dB     \\ 
\hline
\multirow{1}{*}{\textbf{Number of samples}} 
& 1,200,000 &  3,584,000      \\ 
\hline
\multirow{1}{*}{\textbf{Samples distribution across SNR range}} 
& log-uniform distribution & log-normal distribution
\\ 
\hline

\multirow{4}{*}{\textbf{Channel Effects}}
&
\begin{minipage}[t]{\linewidth}\begin{itemize}
\item thermal noise 
\item simple multi-path fading 
\item center-frequency and sample rate offset 
\end{itemize}\end{minipage}
& 
\begin{minipage}[t]{\linewidth}\begin{itemize}
\item path loss
\item 3GPP channel model with correlated slow and fast multipath fading
\item center-frequency and sample rate offset 
\end{itemize}\end{minipage}
\\ 
\midrule \bottomrule[\heavyrulewidth]
\end{tabular}%
}
\end{table}

\section{Transformer Architectures}

The Transformer model, presented in \cite {vaswani2017attention}, marked a significant advancement in the field of sequence-to-sequence processing. Comprising an encoder and a decoder, each with multiple layers, this model 
efficiently maps an input sequence of symbols (words) to a sequence of continuous representations. The decoder then transforms these representations to produce an output sequence. This architecture has demonstrated remarkable accuracy across a variety of sequence-to-sequence tasks, including machine translation and text summarization.

The Transformer architecture, initially designed for language tasks, has been effectively adapted for image processing in the ViT paper \cite{dosovitskiy2020image}. Unlike the original transformer model, which includes both an encoder and decoder for sequence-to-sequence tasks, ViT adapts the Transformer's encoder to process images.  ViT treats an image as a sequence of patches and applies the Transformer encoder to these patches to perform image classification and process images as sequences of flattened patches. The encoder consists of several identical layers, each comprising two sub-layers: a multi-head self-attention mechanism and a feed-forward neural network. Additionally, it has residual connections and layer normalization to mitigate the vanishing gradient issue that arises with deep models. The feed-forward neural network consists of a fully connected layer architecture. In the self-attention mechanism, each element of the input sequence interacts with all other elements to calculate attention weights, highlighting the importance of relationships among different positions within the sequence. The multi-head attention mechanism splits the attention calculation across several heads, with each head independently performing the attention computation. This allows each head to concentrate on distinct features of the input sequence, providing a better understanding of the sequence. This shows how the Transformer can be used not just for text but also for images, demonstrating its versatility.

\begin{figure}[!ht]
\centering
\vspace{0pt}
\includegraphics[trim=270pt 60pt 250pt 80pt, clip, width=6.0in]{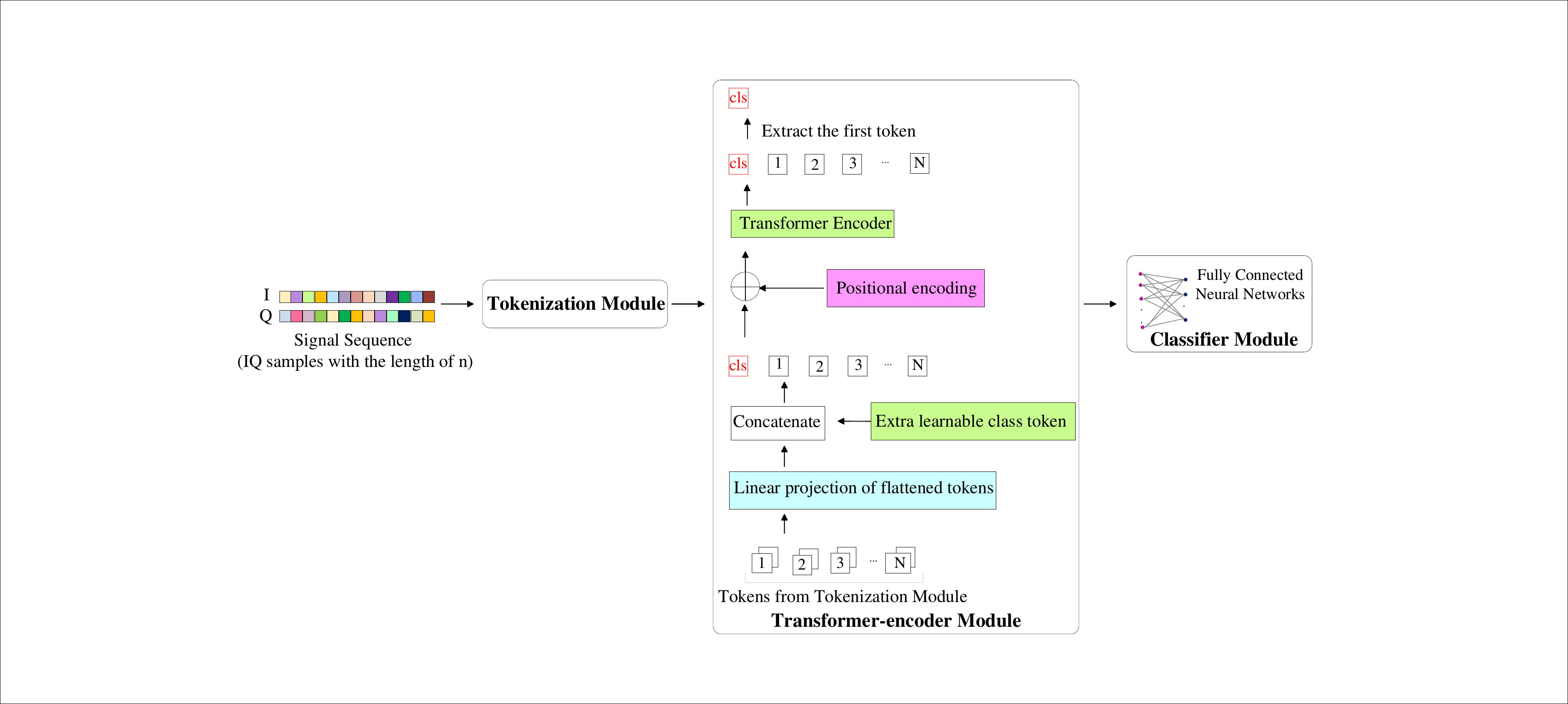}
\caption{The overall architecture of our proposed transformer-based model for the AMR task includes three main components: a Tokenization module that converts the signal into tokens, a Transformer-encoder module, which captures information and extracts relevant features through self-attention mechanism, and a Classifier module for the final classification step.}
\vspace{5pt}
\label{fig_0}
\end{figure}

In this section, we introduce different methods for recognizing signal modulation types using a Transformer-based architecture.  Figure \ref {fig_0} illustrates the overall architecture of our Transformer-based approach, comprising three key components: The Tokenization module, the Transformer-encoder module, and the Classifier module. 
The raw IQ data, composed of in-phase (I) and quadrature (Q) components, forms a two-channel input. However, transformer networks require their inputs to be in the form of tokens, which is achieved through the tokenization module. Various strategies for tokenization are employed in each transformer model discussed in subsequent sections.
In all these architectures, the input undergoes tokenization to generate tokens before being fed into the Transformer-encoder module for capturing relevant features from the data, and the classification task concludes with the output from the classifier module, which is a fully connected neural network.
Utilizing various techniques to generate tokens, we investigate four distinct transformer architectures for the AMR task, namely TransDirect, TransDirect-Overlapping, TransIQ, and TransIQ-Complex. These methods are explained further in the following subsections.

\subsection{TransDirect}
\begin{figure}[!ht]
\centering
\vspace{0pt}
\includegraphics[trim=450pt 20pt 100pt 880pt, clip, width=6in]{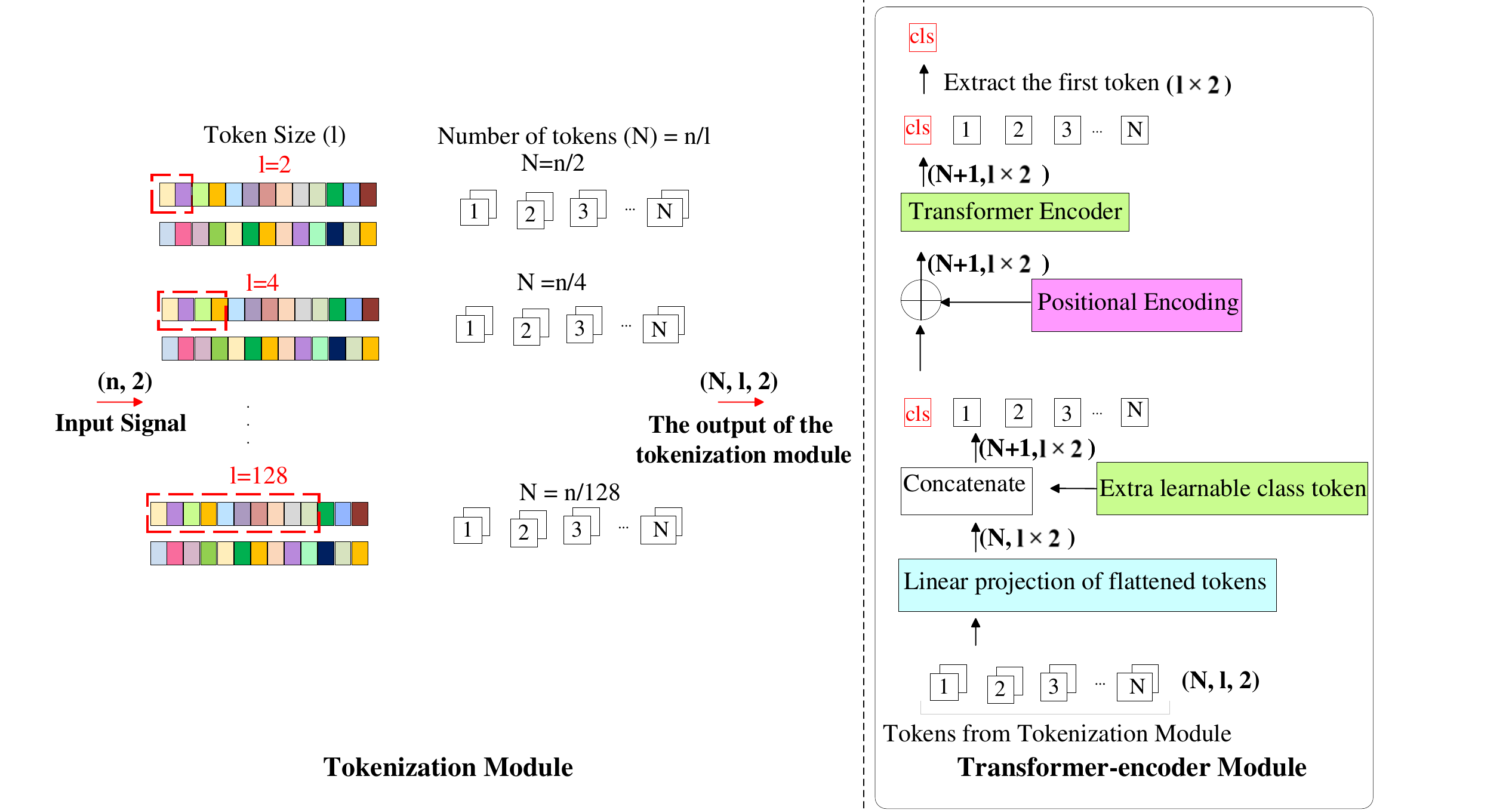}
\caption{TransDirect architecture. In the Tokenization module of this architecture, IQ samples segments into shorter sequences referred to as tokens, each having a size of \(l\).}
\vspace{5pt}
\label{fig_1}
\end{figure}

The detailed architecture considered for TransDirect is shown in Figure \ref {fig_1}. The process starts with the tokenization of data, a crucial step in implementing the Transformer network. This approach is similar to ViT, where image patches are generated from segments of the input as tokens.

In the context of TransDirect, tokenization involves dividing the IQ sample data, which has a sequence length of n and consists of two channels into multiple shorter subsequences that are called tokens. The I and Q samples from the signal sequence can be expressed as  \( I=[i_1, i_2, ..., i_n] \) and  \( Q=[q_1, q_2, ..., q_n] \), with \(n \) denoting the length of the signal. In this arrangement, both I and Q sequences are divided into shorter segments of length \(l\), using a sliding window technique with a step size of \(w\). As the tokens (segments) are designed to be non-overlapping, the stride length \(w\) is set equal to the segment length (token size) l, ensuring distinct and consecutive tokens. Consequently, the segmented subsequences for the I are given by \( I_1=[i_1, i_2, ..., i_l] \), with subsequent sequences like
\( I_2=[i_{1+w}, i_{2+w}, ..., i_{l+w}] \) , leading up to \( I_N=[i_{1+(N-1)w}, i_{2+(N-1)w}, ..., i_{l+(N-1)w}] \). A similar segmentation applies to the Q sequence. As a result, the overall count of tokens using this method is determined by the \( N = \frac{n}{l}\). 

Until this point, the model converts the two-dimensional vector at each time step into \(N \) tokens with two channels. These tokens are then processed through a Transformer-encoder module, according to Figure \ref{fig_0}, where they first get transformed into linear sequences by the Linear Projection Layer, acting as the starting point of the transformer encoder’s processing. Moreover, an additional trainable "class token" is also included with the tokens, increasing the total count of tokens to \(N+1 \). Positional encoding is then applied to the tokens before they are inputted into the Transformer Encoder, enabling the model to understand the relative positioning of the tokens. The tokens then pass through \(N_l \) transformer encoder layers that utilize the self-attention, with a hidden size (embedding size) of \(m\) and attention heads of \(N_{\text{head}}\). As the core part of the transformer, the attention mechanism maps a query and a set of key-value pairs to the output, establishing connections between various positions within a sequence to integrate information across the entire input data. After passing through \(N_l \)  transformer encoder layers, the output is a \((N+1) \times d\) matrix, where d  is the embedded dimension, is calculated as twice the number of samples per token \(l\), reflecting the dual-channel structure of input data.
The first token from this output is then extracted and fed into the classifier module. This module consists of fully connected layers with a set number of hidden layers of \(N_{\text{hidden}}\). The final result of the classification comes from the output of the classifier module, generating a c-dimensional vector, where each represents the probability of a particular modulation type.

\subsection{TransDirect-Overlapping}
\begin{figure}[!ht]
\centering
\vspace{0pt}
\includegraphics[trim=520pt 330pt 510pt 60pt, clip, width=4in]{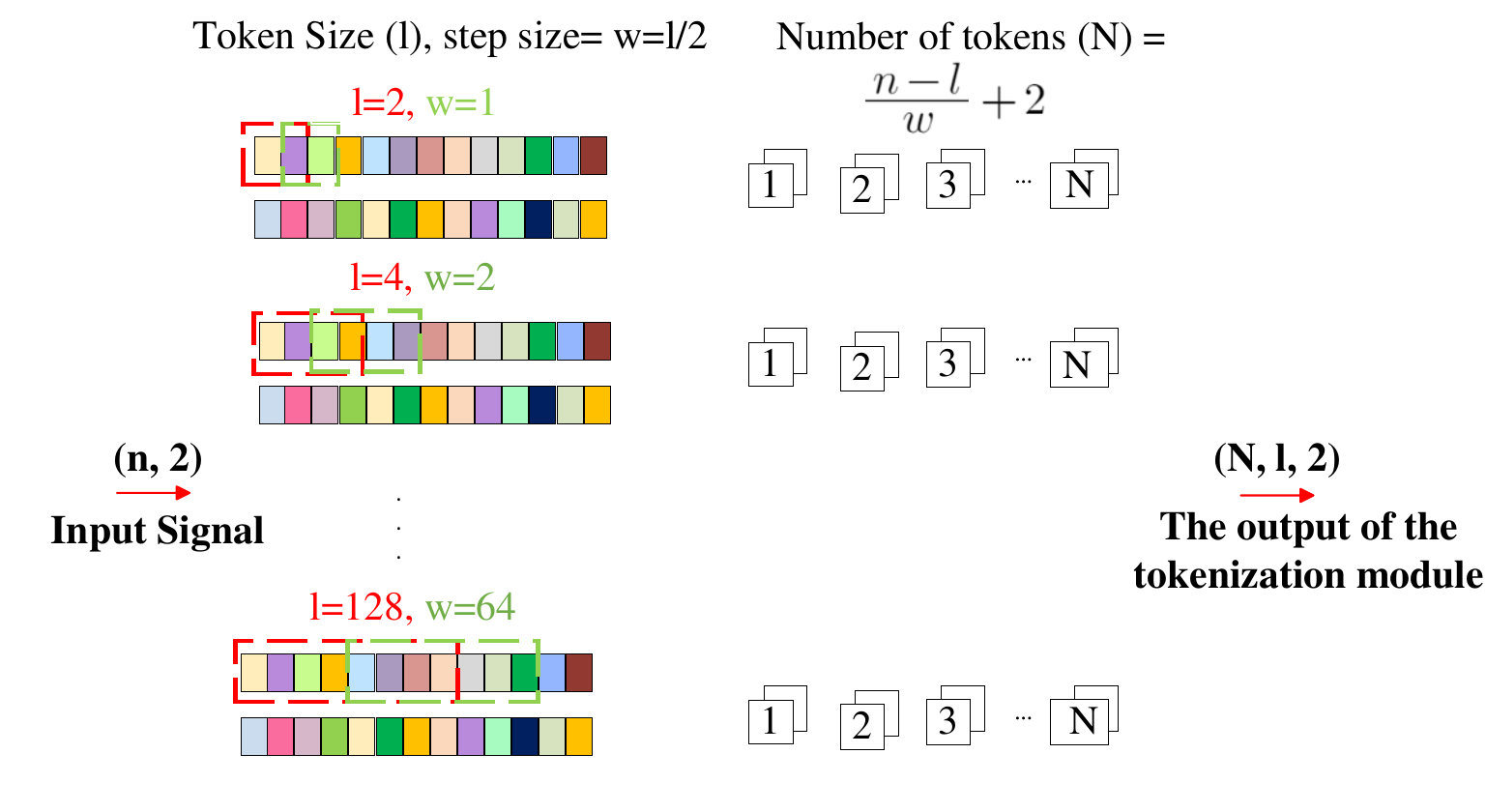}
\caption{Tokenization module of TransDirect-Overlapping architecture, which divides IQ samples into tokens, each with a length \(l\), where each token overlaps the preceding one by \(l/2\).}
\vspace{5pt}
\label{fig_2}
\end{figure}

The overall architecture of the TransDirect-Overlapping is illustrated in Figure \ref{fig_2}. Similar to TransDirect, the original I and Q sequences are segmented into shorter tokens of length $l$ through the use of a sliding window method in the tokenization module. However, unlike the previous method, this technique introduces an overlap between consecutive tokens. Specifically, the step size $w$ is configured to be  \(l/2\), resulting in a 50\% overlap between adjacent tokens. This overlapping strategy allows for a more comprehensive analysis through the reuse of data across tokens. The sub-sequences for the I component are thus initiated with \( I_1=[i_1, i_2, ..., i_l] \), with the following sub-sequences such as \( I_2=[i_{1+(l/2)}, i_{2+(l/2)}, ..., i_{l+(l/2)}] \) and continuing to \( I_N=[i_{1+(N-1)(l/2)}, i_{2+(N-1)(l/2)}, ..., i_{l+(N-1)(l/2)}] \) and this segmentation approach is similarly applied to the Q component. Accordingly, the total number of tokens generated by this method is calculated as \( N=\frac{n-l}{w}+2 \). 
Despite the different tokenization techniques, TransDirect-Overlapping retains the same Transformer-encoder module and Classifier as TransDirect, offering an improved method for initial data segmentation via a sliding window technique that ensures data overlap, thereby maximizing data utilization. Like the TransDirect tokenization approach, the output dimensionality of the tokenization module remains \((N, l, 2)\). However, TransDirect-Overlapping sets itself apart by increasing the number of tokens to \( N=\frac{n-l}{w}+2 \), while maintaining the embedding dimension similar to TransDirect. This dimension equals twice the number of samples per token \(l\), a feature unchanged from TransDirect due to the consistent number of channels (2), presenting I and Q components.

\subsection{TransIQ}
\begin{figure}[!ht]
\centering
\vspace{0pt}
\includegraphics[trim=460pt 860pt 200pt 40pt, clip, width=5.5in]{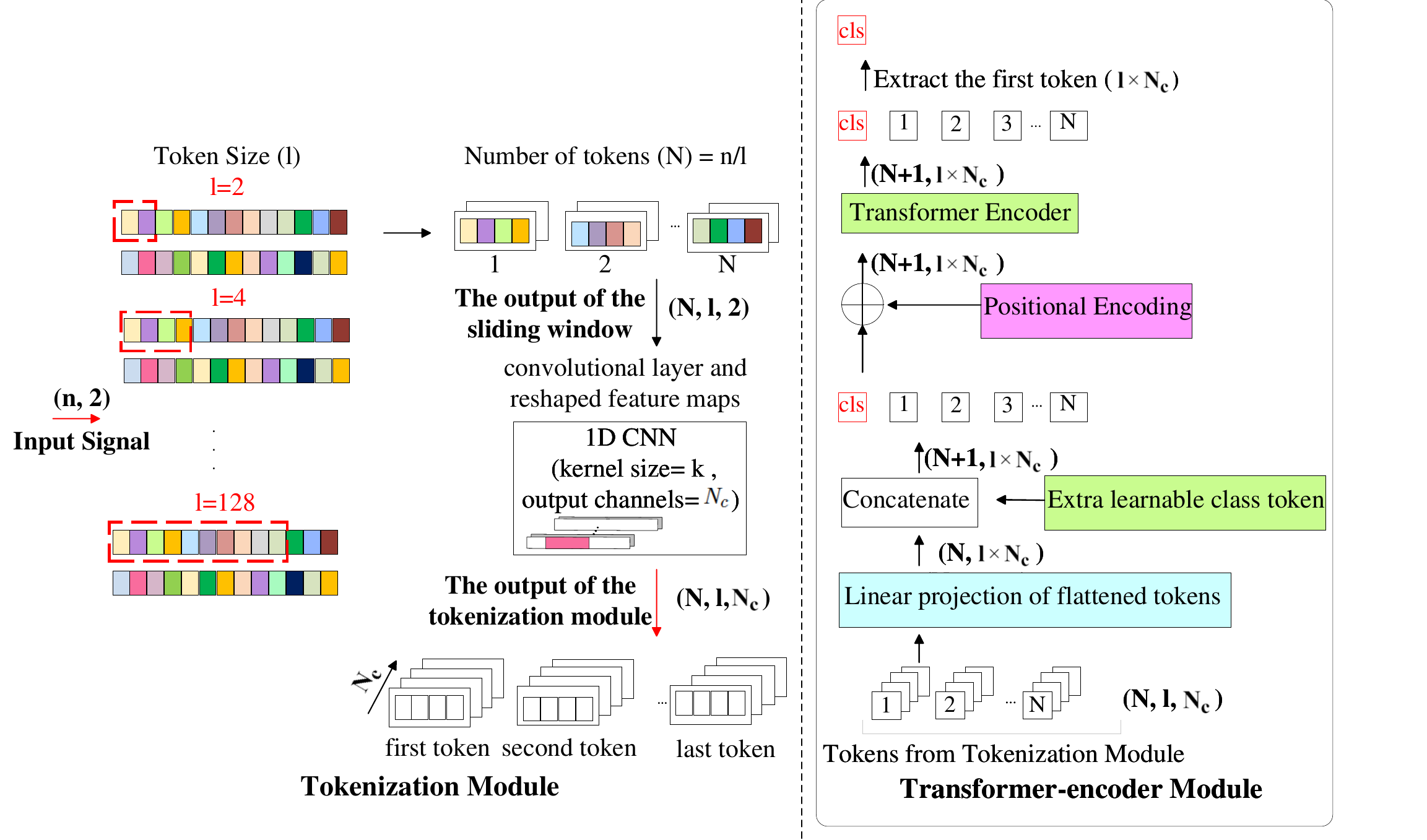}
\caption{TransIQ architecture. In the Tokenization module of this architecture, the input signal segments into tokens. Each token then undergoes one-dimensional convolutional before being processed by the Transformer-encoder module.}
\vspace{5pt}
\label{fig_3}
\end{figure}

In the TransIQ design, as illustrated in the Figure \ref{fig_3}, we enhance the model's feature extraction capacity by expanding its receptive field in the tokenization phase.
Following the tokenization approach of earlier techniques, we first slice the IQ sequence into shorter segments using a sliding window technique in the tokenization module. These segments are then reorganized into a matrix of signal sequences and one-dimensional convolutional encoding is performed.

The introduction of a CNN layer between the tokenization module and the transformer-encoder plays a crucial role in enhancing the feature representation of the input data and facilitates the extraction of detailed features from each token. The CNN layer, through its convolutional filters (kernels), is inherently designed to extract spatial features from the data. Each filter specializes in identifying specific patterns within the input. By applying these filters across the signal sequences, the CNN layer can highlight important features from the input data. This process effectively prepares a rich, detailed feature map for each token. On the other hand, the transformer-encoder utilizes a self-attention mechanism to weigh the importance of different features within the input. To elaborate more, after the CNN layer enriches the feature representation, the Transformer encoder assesses these features, determining their relevance to the task of modulation classification. The self-attention mechanism allows the model to focus on the most relevant features by assigning higher weights to them. This capability stems from the Transformer's design to dynamically allocate attention based on the input's context, enabling it to detect which features (enhanced by the CNN layer) are crucial for understanding the input sequence's characteristics. So, the CNN layer ensures that the Transformer encoder receives a high-resolution feature map, where important characteristics are already highlighted. Subsequently, the Transformer can efficiently evaluate these features' relevance, adjusting its internal weights to prioritize tokens crucial for accurate modulation classification. So, leveraging CNNs for improved feature representation and Transformers to dynamically adjust the importance of those features, TranIQ efficiently allocates suitable weights to each feature segment, focusing on key positions in the input temporal data.

So, the tokens generated from the sliding window are fed into a convolutional layer with a kernel size of k with same-padding and Nc number of output channels. The output of the convolution layer is fed to a ReLU activation function. Up to this stage, the model transforms the two-dimensional vector at each time sample into a \(N_c\) (output channel) dimensional feature representation using different kernels in the convolutional layer. This matrix is then processed through  \(N_l\) transformer encoder layers, which operates on a \((N+1)\times d\) matrix using self-attention mechanism. Here, \(N+1\) represents the total number of tokens, and \(d\) represents the embedding dimension of the transformer. In this scenario, \(d\) equals the product of the token size and the number of output channels of the CNN layer. The first token from the output of the transformer encoder is then fed into the classifier module to perform the modulation classification.

\subsection{TransIQ-Complex}
We have developed a novel strategy for working with complex (I and Q) information in our tokenization module, inspired by the DeepFIR\cite{Restruccia2020} model. This model has shown its effectiveness in AMR tasks by using a complex CNN layer with complex weights and biases. Considering the composite nature of input data, featuring both real and imaginary components, the incorporation of a complex convolutional neural network (CNN) layer within our tokenization module emerges as a key strategy for enhanced data handling. This enhancement aligns with the TransIQ architecture as illustrated in Figure \ref{fig_3}, while integrating two essential modifications.  Primarily, the model processes the input signal as a singular channel entity,  diverging from the conventional dual-channel data approach. Specifically, although the input signal originally consists of two channels, one for the I and another for the Q  components, utilizing a complex convolutional layer allows for the integration of these channels into a unified one-channel input, thereby resizing the signal's dimension to \(n\times 1\). Following initial segmentation into shorter tokens of length \(l\), the dimensionality of each token after applying sliding window is maintained at \(l \times 1\). A second modification is the inclusion of a complex convolutional layer in place of a standard CNN in the architecture shown in Figure \ref{fig_3}. Subsequently, each token undergoes processing via a complex convolutional layer that individually extracts features from the singular-channel data, generating an output with \(N_c\) output channels. Thus, this tokenization method transforms the one-dimensional vector of the signal sequence into a multidimensional feature representation.


\section{Experimental Result and Discussion}

\subsection{Ablation Study}
For evaluating the performance of our proposed models, we utilized the RadioMl2016.10b and our CSPB.ML.2018+ datasets. These datasets were randomly divided, allocating 60\% for training, 20\% for validation, and 20\% for testing. Our experimental setup was based on the PyTorch framework, and we employed the Cross Entropy (CE) loss function. Furthermore, all experiments were performed on a GPU server equipped with 4 Tesla V100 GPUs with 32 GB memory. The Adam optimizer was used for all experiments. Each model underwent training for 100 epochs, starting with a learning rate of 0.001. For experiments involving token sizes larger than 16 samples, a reduced learning rate of 0.0001 was employed with a batch size of 256. Classification accuracy is also evaluated using the F1 score, a key performance metric for classification problems that combines precision and recall into a single measure by calculating their harmonic mean.

In all our experiments, we employed the same structure unless changes from this configuration are specifically mentioned. Our implementation comprises a Transformer encoder architecture with four encoder layers, two attention heads, and a feed-forward network (FFN) with a dimensionality of 64. In the classifier module, we utilize a fully connected neural network consisting of a single hidden layer with 32 neurons with ReLU activation function and dropout, followed by an output layer adjusted to the number of modulation types in each dataset. 

Our investigation began with TransDirect architecture, evaluating the impact of token size on model performance, focusing on how the token length affects classification accuracy. For experiments, we used token lengths of 8, 16, 32, and 64, which corresponds to creating 128, 64, 32, and 16 tokens respectively from the original signal sequence length of 1024 in the CSPB.ML.2018+ dataset. According to our results detailed in Table \ref{tab:tokenization}, we noticed that accuracy increases by about 3\% when the token size doubles from 8 to 16, reaching 56.29\% at its peak and then dropping by about 4\% when the sample size per token is 64. The embedded dimension, in this structure, equals the number of samples per token times the number of channels, which is 2 for I and Q channels. For example, with the TransDirect model using a token size of 16, each token consisting of 16 samples across 2 channels. Subsequently, in the Linear Projection Layer, these tokens are converted to the linear sequence of 32. This leads to an embedding dimension of 32, calculated by multiplying the 16 samples by the 2 channels.
Therefore, as the number of samples per token increases from 8 to 64, while maintaining fixed numbers of encoder layers and head attentions, the total number of parameters grows from 17.2 K to 420 K. This demonstrates that the TransDirect reaches optimal performance when using a token size of 16 samples, understanding the significance of token size and the model's complexity in enhancing classification accuracy while taking into account the computational constraints of IoT devices. This experiment shows a trade-off between the optimal token size and model complexity when tokens are input directly to the transformer encoder.

We conducted additional tests using the TransDirect-Overlapping architecture, where each token overlaps the previous one by 50\%. This variation in tokenization led to a doubling of the number of tokens compared to the initial setup, TransDirect. Specifically, when changing the number of samples per token from 8 to 64, the total number of tokens ranged from 256 (for 8 samples per token) to 32 (for 64 samples per token). Despite this, the total number of model parameters stayed the same as in the TransDirect technique, because the embedding dimension do not change. Table \ref{tab:tokenization} further illustrates a consistent trend in model accuracy, showing an increase of approximately 6\%, from 53.98\% with the token size of 8 to 60.08\% when the token size is increased to 32, and then starts to decrease by approximately 3\% as the token size increases to 64. 

We expanded our investigation into the TransIQ model through a series of comprehensive experiments. We adjusted the network's depth, kernel configurations, and the number of output channels to achieve an optimal balance between accuracy and model size. Starting with the previously implemented Transformer-encoder configuration of 2 heads and 4 layers, we explored token sizes ranging from 8 to 32, integrating a convolutional layer with 8 output channels. The highest accuracy we achieved was 63.72\%, with the optimal token size set to 16 samples. This represents an approximately 7\% improvement in accuracy compared to the best performance of the TransDirect and an approximately 3\% increase over the TransDirect-Overlapping model. The improved performance is due to the inclusion of a convolutional layer, which enhances the model's capability to extract features from each token, distinguishing it from the TransIQ and TransIQ-Overlapping models.

In another set of experiments, we opted for a token size of 8 to constrain the model's size. We then explored different configurations for head and layer within our TransIQ, seeking to enhance efficiency by minimizing the parameter count. As the total number of parameters is proportional to the CNN complexity and embedding dimension of the Transformer encoder, we evaluated two versions: the first variant, named the Large Variant of TransIQ, features a convolutional layer with 8 output channels, 4 heads, and 8 layers, achieving the highest accuracy on the dataset at 65.80\%. The Small Variant of TransIQ, on the other hand, utilizes a token size of 8 and a Transformer-encoder architecture with 2 heads and 6 layers. This model has a total of 179 K parameters and achieves an accuracy of 65.39\%.

We further explored the impact of integrating a complex convolutional layer by employing the complex layer in the TransIQ-Complex model. This model, distinct from typical CNNs, incorporates complex weights and biases in its CNN layer. As the token size expanded from 8 to 16, 
we observed an accuracy improvement of approximately 1.6\%. However, considering the increase in model complexity and the marginal gains in accuracy, the use of a complex layer proved to be minimally beneficial in scenarios where model size is a critical consideration.

\begin{table}[htpb]  
\centering
\caption{A comparative analysis of different methods of tokenization in terms of F1 score and number of parameters on CSPB.ML.2018+ dataset.}
\label{tab:tokenization}
\resizebox{\columnwidth}{!}{%
\begin{tabular}{l|c|ccc}
\toprule[\heavyrulewidth] \midrule
&  \textbf{Tokenization}   & \begin{tabular}[c]{@{}c@{}}\textbf{F1 score}\\ \end{tabular}  & \begin{tabular}[c]{@{}c@{}}\textbf{Number of parameters}\\\end{tabular}   \\ \midrule
\multirow{4}{*}{\textbf{TransDirect}} 

& 8 samples   & 53.15  & 17.2 K\\
& 16 samples      & \textbf{56.29} & 44.1 K  \\
 & 32 samples     & 56.20 & 128 K \\ 
 & 64 samples     &  52.24  & 420 K \\ 
\hline
\multirow{4}{*}{\textbf{TransDirect-Overlapping }}
 & 8 samples  &      53.98      & 17.2 K \\ 
 & 16 samples  &    59.43 &  44.1 K  \\ 
 & 32 samples   &   \textbf{60.08}  &  128 K  \\
 & 64 samples    &  57.67    &420 K  \\

 \hline
\multirow{7}{*}{\textbf{TransIQ}} 
 & 8 samples with 8 output channels (head=2, layer=4)          &  63.69  & 128 K\\
 & 8 samples with 16 output channels (head=2, layer=4)          & 63.25  &420 K\\
 & 16 samples with 8 output channels (head=2, layer=4)    &  63.72  &420 K \\ 
 & 32 samples with 8 output channels (head=2, layer=4)     &   62.68   & 1.5 M   \\ 

 & 8 samples with 8 output channels (head=4, layer=8)           &  \textbf{65.80} & 229 K \\

& 8 samples with 8 output channels (head=2, layer=6)   & \textbf{65.39} & 179 K \\
 \hline
 \multirow{2}{*}{\textbf{TransIQ-Complex}} 
& 8 samples with 8 output channels (head=2, layer=4)      &   61.19  &  420 K \\
& 16 samples with 8 output channels (head=2, layer=4)   &  \textbf{62.79}  & 1.5 M
 \\ 
 & 32 samples with 8 output channels (head=2, layer=4)  &   59.33    &    5.6 M    \\ 
\midrule \bottomrule[\heavyrulewidth]
\end{tabular}%
}
\end{table}

Referencing Table \ref{tab:tokenization}, an analysis comparing the F1 score and parameter count across different token sizes (8 to 64) for four methods of tokenization reveals detailed insights. A minimal increase in accuracy, a 0.83\% improvement, is observed when comparing TransDirect to TransDirect-Overlapping for a token size of 8, but there is a more significant improvement, nearly 10\%, from TransDirect-Overlapping to the TransIQ model. For the token size of 16, both the TransDirect and TranDirect-Overlapping architectures have the same number of parameters of 44.1 K, but TransDirect-Overlapping achieves an approximately 3\% higher accuracy due to its overlapping tokenization technique. Furthermore, adopting the TransIQ model with the same token size, and with a  convolutional layer with 8 output channels, increases the parameter count to 420 K. This change results in a 4\% accuracy improvement over TransDirect-Overlapping and an impressive 7\% improvement compared to TransDirect with the same token size. Conversely, the TransIQ-Complex architecture, despite increasing its parameters to 1.5 M with the token size of 16, experiences a roughly 1\% drop in accuracy compared to the TransIQ with the same token size. This indicates that while token size plays a crucial role in achieving higher accuracy, simply increasing model complexity does not assure improved performance. Identifying the optimal balance requires thorough experimentation, as evidenced by the data presented in Table \ref{tab:tokenization}. 


\subsection{Comparison with other baseline methods}
To evaluate the performance of our proposed methods, we conducted a quantitative analysis comparing it against models with varying structures on two different datasets, including RadioML2016.10b and CSPB.ML.2018+. Our initial analysis focused on RadioML2016.10b dataset, evaluating models including DenseNet\cite{liuDeepNeuralNetwork2017a}, CNN2 \cite{tekbiyikRobustFastAutomatic2020}, VTCNN2\cite{hauser2017signal}, ResNet\cite{o2018over}, CLDNN\cite{liuDeepNeuralNetwork2017a}, and Mcformer\cite{hamidi2021mcformer}. Among these models, DenseNet, CNN2, VTCNN2, and ResNet are built upon CNN, and CLDNN integrates both RNN and CNN architectures. In this comparison, we also included another baseline model, Mcformer, which is built on the Transformer architecture.  However, since the Mcformer model was not developed using the PyTorch framework and our inability to access its detailed architecture, we were unable to replicate the results and determine the number of parameters. Therefore, we referenced the results for Mcformer from \cite {hamidi2021mcformer}. 

In the initial comparison, we tested two variants of our TransIQ model against baselines on the RadioML2016.10b dataset. The TransIQ-Large Variant featured a token size of 8 samples, a convolutional layer with 8-output channels, and a transformer encoder with 4 heads and 8 layers, and the TransIQ-Small Variant, with the same token size and convolutional layer but a transformer encoder with 2 heads and 6 layers. 

The experimental results are shown in Tables \ref {tab:F1_1}. As demonstrated in this table, our proposed model outperforms all baseline models in terms of accuracy on this dataset. The table illustrates that our proposed models achieve roughly a 9\% better accuracy compared to the DenseNet model, despite having 14 times fewer parameters in the TransIQ-Large Variant and 18 times fewer in the TransIQ-Small Variant. Moreover, both TransIQ model variants demonstrate superior accuracy to ResNet, with only a minimal increment in the number of parameters, making their parameter counts relatively comparable. Our analysis, summarized in Table \ref {tab:F1_1}, underscores the TransIQ architecture's impressive capability in the AMR task, combining high accuracy with reduced parameter needs. To further validate the robustness and potential of our model, we expanded our evaluation to include another dataset characterized by more channel effects, aiming to test our model's capability under more challenging scenarios. 

Consequently, we assessed the performance of our proposed models on the CSPB.ML.2018+ dataset, which has signals of longer length and more channel effects compared to the RadioML2016.10b dataset. The results, detailed in Table \ref{tab:F1_2}, show that while the parameter counts for baseline models increase significantly with this dataset, the complexity of our proposed models remains consistent. This consistency in complexity is attributed to the same embedding dimension. This demonstrates our model's adaptability and efficiency, maintaining their complexity unchanged even when applied to datasets featuring signals of longer lengths.

\begin{table}[!ht]
\centering
\caption{Comparative Analysis of F1 Scores and Parameter Counts Across Various Models on the RadioML2016\cite{o2016radio} dataset.}
\label{tab:F1_1}
 \small{
\begin{tabular}{l|cccc}
\toprule[\heavyrulewidth] \midrule
\textbf{Methods}     & \textbf{F1 Score} & \textbf{Number of Parameters}\\ \midrule
\textbf{DenseNet \cite{liuDeepNeuralNetwork2017a}}        & 56.93     &  3.3 M   \\
\textbf{CLDNN \cite{liuDeepNeuralNetwork2017a}}        &  61.14    &    1.3 M    \\
\textbf{VTCNN2\cite{hauser2017signal}}        &61.53        &    5.5 M    \\
\textbf{CNN2 \cite{tekbiyikRobustFastAutomatic2020}}   &   60.94         &  1 M   \\
\textbf{ResNet \cite{o2018over}}   &     64.62          &  107 K   \\
\textbf{Mcformer \cite{hamidi2021mcformer}}    & \textbf{65.03}       &   -           \\ \midrule


\textbf{TransIQ (Large variant)}     &  \textbf{65.75} &  229 K\\
\textbf{TransIQ (Small variant)}    & \textbf{65.61} &  179 K\\

\midrule \bottomrule[\heavyrulewidth]
\end{tabular}%
}
\end{table}

\begin{figure}[!ht]
\centering
\vspace{-1pt}
\includegraphics[trim=0pt 0pt 0pt 30pt, clip, width=5.0in]{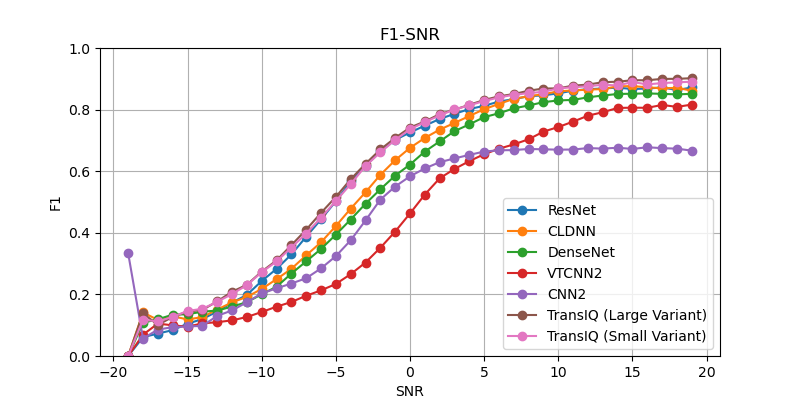}
\caption{Modulation recognition accuracy comparison between two variant of TransIQ and other baseline models on CSPB.ML.2018+ dataset with a change in SNR, where SNR ranged from -19 dB to +20 dB.}
\label{fig_DL2}
\end{figure}

Figure \ref{fig_DL2} shows how accuracy changes for different methods on the CSPB.ML.2018+ dataset when the SNR changes. It is observed that classification accuracy increases with higher SNR levels across all neural network architectures. We note that the larger variant of TransIQ consistently outperforms the baseline models across all SNR values. The small variant of TransIQ also shows competitive performance against the baselines. In particular, the smaller variant has 179 K parameters, significantly less than the millions of parameters found in all baseline models except ResNet. Remarkably, the large variant of TransIQ achieves better accuracy than the ResNet model while they are comparatively similar in terms of the number of parameters. The superiority of our model becomes particularly apparent, especially in situations with low SNR (\(SNR <0 \)). Although the TransIQ-Large Variant outperforms the Small one, this performance improvement comes with the increased computational complexity of 50 K. However, both architectures have significantly fewer parameters compared to the other baselines. This highlights the suitability of our proposed models for IoT applications.

\begin{table}[!ht]
\centering
\caption{Comparative Analysis of F1 Scores and Parameter Counts Across Various Models on the CSPB.ML.2018+ dataset.}
\label{tab:F1_2}
 \small{
\begin{tabular}{l|cccc}
\toprule[\heavyrulewidth] \midrule
\textbf{Methods}      & \textbf{F1 Score } & \textbf{Number of Parameters}\\ \midrule
\textbf{DenseNet \cite{liuDeepNeuralNetwork2017a}}            &  57.87   &21.6M  \\
\textbf{CLDNN \cite{liuDeepNeuralNetwork2017a}}           &     61.26  &7.1M  \\
\textbf{VTCNN2\cite{hauser2017signal}}              &      47.29  & 42.2M \\
\textbf{CNN2 \cite{tekbiyikRobustFastAutomatic2020}}          &  52.57  &3.8M \\
\textbf{ResNet \cite{o2018over}}        & 65.48  & 164 K   \\ \midrule

\textbf{TransIQ (Large variant)} &  65.80       &  229 K\\
\textbf{TransIQ (Small variant)} & 65.39       &  179 K\\
\midrule \bottomrule[\heavyrulewidth]
\end{tabular}%
}
\end{table}

\label{fig_DL4}
\label{fig_DL4}

The confusion matrix, illustrated in Figure \ref{fig_DL4}, shows the performance of the top-performing baseline model (ResNet) as well as TransIQ variants on the CSPB.ML.2018+ dataset. This matrix is a critical tool for evaluating model accuracy, detailing the relationship between actual and predicted class instances through its structured design. It is a \(N * N \) matrix, where \(N\) represents the number of classes for classification. Specifically, each row in the matrix corresponds to actual class instances, while each column reflects the model's predictions. The numbers along the diagonal of the matrix represent the true positives for each class, meaning the number of times the model correctly predicted each class. On the other hand, off-diagonal elements represent misclassifications. According to Figure \ref{fig_DL4}, here the confusion matrix is an \(8 * 8 \) matrix, aligning with the 8 different modulation types in the CSPB.ML.2018+ dataset. For instance, in the confusion matrix of the ResNet model, the value located at the intersection of the first row and the first column is 0.97. This indicates that 97\% of the instances truly belonging to BPSK modulation were correctly classified as BPSK by the model. To illustrate, the element at the second row and first column of ResNet's confusion matrix has a value of 0.02, indicating that 2\% of the instances that are actually QPSK modulation were incorrectly predicted as BPSK by the model.

Further analysis of it reveals that ResNet still has significant confusion between QAM modulations. Some higher-order PSK modulations are also classified as QAM. TransIQ is better able to correctly classify PSK modulations and shows a small amount of confusion among higher-order PSK. TransIQ is also much better able to discern between 16QAM and other higher-order QAM modulations. Both variants of TransIQ showed similar confusion between 64- and 256-QAM.

\begin{figure}[!ht]
\centering
\vspace{-1pt}
\includegraphics[trim=0pt 0pt 0pt 0pt, clip, width=6.0in]{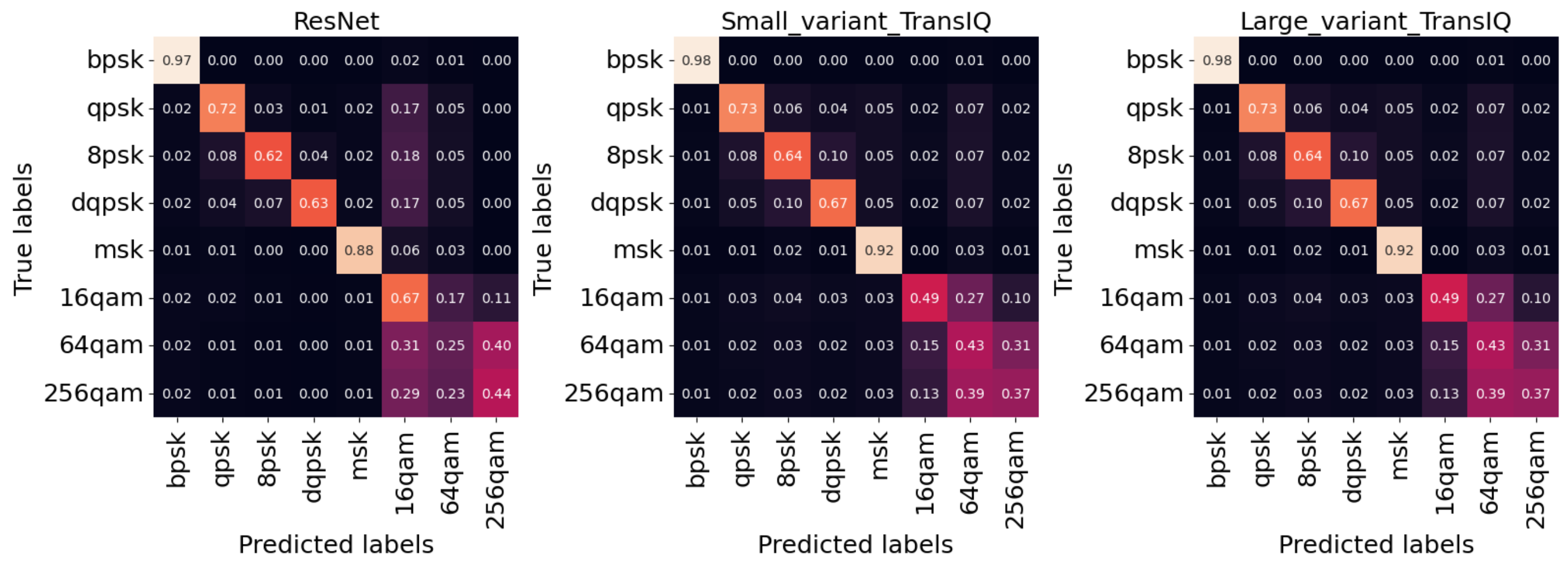}
\caption{Confusion matrix for ResNet versus proposed models (TransIQ-Large Variant and TransIQ-Small Variant) on CSPB.ML.2018+ dataset. Each row represents the actual class, and each column corresponds to the class as predicted by the mentioned models.}
\label{fig_DL4}
\end{figure}

\subsection{Latency and Throughput Metrics}
In assessing the performance of TransIQ (Large and Small variant), two metrics of latency and throughput are analyzed as well. These analyses offer insights into the models' effectiveness and suitability for real-world scenarios. Latency is defined as the duration required for the model to process a single input, measured in seconds. On the other hand, throughput, refers to the model's capability to process a certain number of inputs per second.

Our experiments to measure latency and throughput were carried out on the NVIDIA Jetson AGX Orin Developer Kit. This kit is equipped with Ampere architecture GPU with 2048 CUDA cores and is also powered by an ARM-based CPU with 12 cores. The system includes the memory of 64GB, providing sufficient resources for demanding AI tasks. To carry out our measurements, we utilized the CPU resources provided by the kit, in the MAXN performance mode and on NVDIA JetPack version 6.0-b52. 
The measured latency and throughput for two variants of TransIQ are presented in Table \ref{tab:F1_3}. The table illustrates that the latency for TransIQ (Small variant) is 3.36 ms/sample, while the Large variant has a higher latency of 5.93 ms/sample. This increased latency for the Large variant is anticipated due to its greater complexity, with 229k parameters compared to the 179k parameters of the Small variant, requiring more time to make predictions. In terms of throughput, the Small variant demonstrates a throughput of approximately 297 samples per second. This high rate indicates its capability to manage large volumes of predictions with minimal latency, making it well-suited for real-time applications with a bit lower accuracy. Conversely, the Large variant has a throughput of approximately 168 samples per second. Although this is lower compared to the Small variant, the Large variant is suitable for AMR applications requiring higher prediction accuracy, at the cost of slower processing input data.

\begin{table}[!ht]
\centering
\caption{Measured latency and throughput of two variants of TransIQ on NVIDIA Jetson AGX Orin Developer Kit.}
\label{tab:F1_3}
 \small{
\begin{tabular}{l|cccc}
\toprule[\heavyrulewidth] \midrule
\textbf{Models}      & \textbf{Latency (ms/sample) } & \textbf{Throughput (sample/sec) }\\ \midrule
\textbf{TransIQ (Small variant)}           & 3.36    & 297.61 \\
\textbf{TransIQ (Large variant)}           & 5.93    & 168.63 \\
\midrule \bottomrule[\heavyrulewidth]
\end{tabular}%
}
\end{table}

\section{Future Investigation}

While this paper provides encouraging results for transformer-based architecture (TransIQ), there are many possible future research directions. We introduced a new approach to process the signal data that comes in IQ samples. This opens the door to further research on how to represent signals differently. In the future, we plan to combine various methods of  signals representation, including amplitude/phase, and time-frequency. This combination could give us more detailed information and help improve how well we can recognize and understand these signals.

A key area for future research will also be to explore strategies for maintaining high recognition accuracy even with limited data samples. Moving forward, we aim to delve into the development of a transformer encoder designed to achieve high accuracy in scenarios characterized by small sample sizes. This direction not only addresses the challenge of data scarcity but also enhances the method's applicability across various contexts where collecting large datasets is impractical.  
Additionally, the model's robustness may diminish due to the absence of crucial labels. Therefore, exploring semi-supervised or unsupervised learning methods presents a promising avenue for future research. These approaches could potentially enhance the model's ability to learn from limited or unlabeled data, thereby improving its overall effectiveness and reliability.

In future work, we plan to explore the application of transformer-based architectures within distributed environments, especially in scenarios where IoT devices are widely dispersed. Investigating the development and implementation of transformers in such environments represents a significant avenue for advancing the application of transformer techniques in IoT networks. An important aspect of this research will focus on handling multi-receiver datasets, where each receiver is subject to its unique channel effects. This direction aims to address the challenges of integrating sophisticated deep learning models like transformers with the complex dynamics of distributed IoT systems, enhancing their ability to efficiently process and analyze data across varied network conditions.

\section{Conclusions}

In wireless communication, AMR plays a crucial role in efficient signal demodulation. Traditional AMR methods face challenges with complexity and high computational demands. Recent advancements in deep learning have offered promising solutions to these challenges. 
This study highlights the adaptability of the Transformer architecture, originally developed for language processing, to the field of signal modulation recognition. By adapting the Transformer's encoder with various tokenization strategies, ranging from non-overlapping and overlapping segmentation to the integration of convolutional layers, this research introduces a novel architecture, TransIQ, for AMR tasks. A thorough examination of the TransIQ model, utilizing RadioML2016.10b and CSPB.ML.2018+ datasets, was presented. Moreover, the experiments underline the necessity of balancing model size and performance, especially for deployment in resource-constrained environments like IoT devices. Comparative analysis with baseline models shows that TransIQ exhibits superiority in terms of accuracy and parameter efficiency.

\bibliographystyle{unsrtnat}
\bibliography{references}  






\end{document}